# Path-integral evolution of chaos embedded in noise: Duffing neocortical analog

Lester Ingber

*Lester Ingber Research, P.O. Box 857, McLean, Virginia 22101 U.S.A.*

ingber@alumni.caltech.edu

Ramesh Srinivasan

*Department of Psychology and Institute of Cognitive and Decision Sciences,*

*and Electrical Geodesics, Inc., Eugene, OR 97403, U.S.A.*

*ramesh@oregon.uoregon.edu*

*Paul L. Nunez*

*Department of Biomedical Engineering, Tulane University, New Orleans, Louisiana 70118 U.S.A.*

pnunez@mailhost.tcs.tulane.edu

**Abstract**—A two dimensional time-dependent Duffing oscillator model of macroscopic neocortex exhibits chaos for some ranges of parameters. We embed this model in moderate noise, typical of the context presented in real neocortex, using PATHINT, a non-Monte-Carlo path-integral algorithm that is particularly adept in handling nonlinear Fokker-Planck systems. This approach shows promise to investigate whether chaos in neocortex, as predicted by such models, can survive in noisy contexts.

# 1. INTRODUCTION

A global theory of neocortical dynamic function at macroscopic scales was developed to explain salient features of electroencephalographic(EEG) data recorded from human scalp [1]. In some limiting cases, the theory predicts EEG standing waves composed of linear combinations of spatial modes. These modes (which may be called order parameters in the parlance of modern dynamical methods) are governed by linear or quasilinear ordinary differential equations [1-3]. These data and theoretical works suggest several connections to reports of chaotic attractors estimated for EEG data [1]. In order to illuminate possible relations between sub-macroscopic chaos (for example in neocortical columns) and large scale data observed at the scalp, we suggested a simple metaphoric system, the linear stretched string with nonlinear attached springs [4-6]. We have studied a forced partial differential equation describing the string-spring system, where the forcing term might represent a steady-state evoked potential in the neocortical analogy. In the limiting case when string tension is zero, the forced Duffing equation is obtained. The Duffing equation exhibits a rich dynamic behavior of periodic and chaotic motion, for different parameters (associated with different wave numbers in the string-spring system) [7].

Another series of papers has outlined a statistical mechanics of neocortical interactions (SMNI), demonstrating that the statistics of synaptic and neuronal interactions develops a stochastic background of moderate noise that survives up to the macroscopic scales that are typically measured by scalp EEG [4,8-12].

This study is concerned with the relation between chaos that may occur in particular spatial modes (the order parameters) and EEG signals consisting of combinations of modes and noise. In this context, we may consider the "noise" to be either instrumental or biologic, e.g., as detailed in the SMNI papers. That is, can we reasonably expect correlation dimension estimates of EEG data to provide information that can be used to construct physiologically-based theories of neocortical dynamics.

Section 2 describes the Duffing metaphor of neocortex that exhibits chaos. Section 3 describes the embedding of the deterministic model into a stochastic medium. Section 4 describes the PATHINT code used for the stochastic model. Section 5 gives the results of the PATHINT calculations. Section 6 is a brief conclusion.

## 2. DUFFING ANALOG OF NEOCORTEX

At this stage of our understanding, choosing a specific fit to a macroscopic EEG model is necessarily limited to the specific data and would necessarily be controversial. However, a relatively simple physical system in which the general issues of spatial scale interaction and connection to experimental EEG can be discussed with minimal complication is a linear stretched string with attached nonlinear springs [4].

The ends of the string may be considered fixed or the string may form a closed loop to make the system more analogous to the closed neocortical/corticocortical fiber surface. A closed loop of transmission line with nonlinear dielectric conductance is the equivalent electric system. In the proposed metaphorical link with the brain, the springs represent local circuit dynamics at columnar scales whereas the string is believed to be somewhat more analogous to neocortical interactions along the corticocortical fibers [5,10].

String displacement, considered to be somewhat analogous to the modulation of cortical surface potential is governed by

$$\frac{\partial^2 \Phi}{\partial t^2} - c^2 \frac{\partial^2 \Phi}{\partial x^2} + \alpha \frac{\partial \Phi}{\partial t} + [\omega_0^2 + f(\Phi)]\Phi = F(x, t) \ , \tag{1}$$

In the case of fixed-end supports, the forcing function takes the form

$$F(x, t) = B(x) \cos t \ . \tag{2}$$

Here $c^2$ is the usual string tension parameter (or wave velocity parameter), $\omega_0^2$ is proportional to the linear spring constant, and $\alpha$ is the damping parameter. In the neocortical metaphor, $F(x, t)$ might represent ongoing random subcortical input (e.g., from the thalamus), or sensory information designed to study linear/nonlinear resonance. Since the string/springs is simply a metaphor for the brain, the nonlinearity is arbitrary, so we choose a cubic nonlinearity and sinusoidal driving so that string displacement is governed by the partial differential equation

$$\frac{\partial^2 \Phi}{\partial t^2} - c^2 \frac{\partial^2 \Phi}{\partial x^2} + \alpha \frac{\partial \Phi}{\partial t} + \omega_0^2 \Phi + \Phi^3 = B(x) \cos t \ . \tag{3}$$

The driving has been included in order to connect the system to the well studied forced Duffing's equation as well as to obtain chaotic dynamics. In the limiting case of a completely relaxed string (i.e., the velocity of propagation $c^2$ is zero), the springs behave independently and drive the string, governed by the forced

Duffing's equation at $x = x_0$,

$$\frac{\partial^2 \Phi}{\partial t^2} + \alpha \frac{\partial \Phi}{\partial t} + \omega_0^2 \Phi + \Phi^3 = B(x) \cos t \ . \tag{4}$$

The alternative way to arrive at Duffing's equation is to do a spatial mode expansion of the string equation and then only keep the lowest wavenumber [6]. For instance, the Lorenz system is derived in this way from the atmospheric turbulence equations [13].

This equation has been studied numerically for the case $\omega_0 = 0$ in the parameter space $(\alpha, B)$ using spectral analysis and phase-plane projections [7]. A rich structure of periodic, quasi-periodic and chaotic motion were reported. We duplicated Ueda's results using a standard fifth and sixth order Runge-Kutta algorithm (IMSL routine DIVPRK). The dynamics were classified as chaotic on the basis of characteristic broadening of spectral lines and aperiodic trajectories in phase space as well as correlation dimension analysis. In all simulations discussed here, the damping parameter $\alpha = 0.05$ and $B = 5.5$.

Figs. 1a and 1b illustrate the evolution of the phase space for the non-chaos and chaos cases of $\omega_0 = 1.0$ and $\omega_0 = 0.1$ cases, resp., after transients have been passed ($t > 250$). Figs. 2a and 2b illustrate the evolution of the phase space for the non-chaos and chaos cases of $\omega_0 = 1.0$ and $\omega_0 = 0.1$ cases, resp., during a transient period between $t = 12$ and $t = 15.5$. The non-chaos cases are similar in appearance except for the settling of transients.

## 3. EMBEDDING DETERMINISTIC DUFFING SYSTEM IN NOISE

The deterministic Duffing system described above can be written in the form

$$\ddot{x} = f(x, t) \ ,$$

$$f = -\alpha \dot{x} - \omega_0^2 x + B \cos t \ , \tag{5}$$

and recast as

$$\dot{x} = y \ ,$$

$$\dot{y} = f(x, t) \ ,$$

$$f = -\alpha y - \omega_0^2 x + B \cos t \ . \tag{6}$$

Note that this is equivalent to a 3-dimensional autonomous set of equations, e.g., replacing $\cos t$ by $\cos z$,

defining $\dot{z} = \beta$, and taking $\beta$ to be some constant.

We now add independent Gaussian-Markovian ("white") noise to both $\dot{x}$ and $\dot{y}$, $\eta_i^j$, where the variables are represented by $i = \{x, y\}$ and the noise terms are represented by $j = \{1, 2\}$,

$$\dot{x} = y + \hat{g}_x^1 \eta_1 \ ,$$

$$\dot{y} = f(x, t) + \hat{g}_y^2 \eta_2 \ ,$$

$$< \eta^j(t) >_\eta = 0 \ ,$$

$$< \eta^j(t), \eta^{j'}(t') >_\eta = \delta^{jj'} \delta(t - t') \ . \tag{7}$$

In this study, we take moderate noise and simply set $\hat{g}_i^j = 1.0 \delta_i^j$.

The above defines a set of Langevin rate-equations, which can be recast into equivalent Fokker-Planck and path-integral equations [14]. A compact derivation has been given in several papers [10,15]. The equivalent short-time conditional probability distribution $P$, in terms of its Lagrangian $L$, corresponding to these Langevin rate-equations is

$$P[x, y; t + \Delta t | x, y, t] = \frac{1}{(2\pi \Delta t)(\hat{g}^{11} \hat{g}^{22})^2} \exp(-L\Delta t) \ ,$$

$$L = \frac{(\dot{x} - y)^2}{2(\hat{g}^{11})^2} + \frac{(\dot{y} - f)^2}{2(\hat{g}^{22})^2} \ . \tag{8}$$

## 4. PATHINT CODE

PATHINT is a non-Monte-Carlo histogram C-code developed to evolve an $n$-dimensional system (subject to machine constraints), based on a generalization of an algorithm demonstrated by Wehner and Wolfer to be extremely robust for nonlinear Lagrangians with moderate noise [16-18]. This algorithm has been used for studies in neuroscience [19,20], financial markets [21,22], combat analysis [23,24], and in a study of stochastic chaos [25].

PATHINT computes the path-integral of a system in terms of its Lagrangian $L$,

$$P[q_t | q_{t_0}] dq(t) = \int \cdots \int \mathcal{D}q \exp\left(-\min \int_{t_0}^t dt' L\right) \delta(q(t_0) = q_0) \ \delta(q(t) = q_t) \ ,$$

$$Dq = \lim_{u \to \infty} \prod_{\rho=1}^{u+1} g^{1/2} \prod_i (2\pi\Delta t)^{-1/2} dq^i_\rho \ ,$$

$$L(\dot{q}^i, q^i, t) = \frac{1}{2} (\dot{q}^i - g^i) g_{ii'} (\dot{q}^{i'} - g^{i'}) \ ,$$

$$g_{ii'} = (g^{ii'})^{-1} \ ,$$

$$g = \det(g_{ii'}) \ . \tag{9}$$

If the diffusions are not constant, then there are additional terms [14].

The histogram procedure recognizes that the distribution can be numerically approximated to a high degree of accuracy as sums of rectangles at points $q^i$ of height $P_i$ and width $\Delta q^i$. For convenience, just consider a one-dimensional system. The above path-integral representation can be rewritten, for each of its intermediate integrals, as

$$P(x; t+\Delta t) = \int dx' [g^{1/2} (2\pi\Delta t)^{-1/2} \exp(-L\,\Delta t)] P(x'; t)$$

$$= \int dx' G(x, x'; \Delta t) P(x'; t) \ ,$$

$$P(x; t) = \sum_{i=1}^{N} \pi(x - x^i) P_i(t) \ ,$$

$$\pi(x - x^i) = \begin{cases} 1 \ , & (x^i - \frac{1}{2}\Delta x^{i-1}) \leq x \leq (x^i + \frac{1}{2}\Delta x^i) \ , \\ 0 \ , & \text{otherwise} \ . \end{cases} \tag{10}$$

This yields

$$P_i(t+\Delta t) = T_{ij}(\Delta t) P_j(t) \ ,$$

$$T_{ij}(\Delta t) = \frac{2}{\Delta x^{i-1} + \Delta x^i} \int_{x^i - \Delta x^{i-1}/2}^{x^i + \Delta x^i/2} dx \int_{x^j - \Delta x^{j-1}/2}^{x^j + \Delta x^j/2} dx' G(x, x'; \Delta t) \ . \tag{11}$$

$T_{ij}$ is a banded matrix representing the Gaussian nature of the short-time probability centered about the (possibly time-dependent) drift.

Here, this histogram procedure s used in two dimensions, i.e., using a matrix $T_{ijkl}$ [23]. Explicit dependence of $L$ on time $t$ also can be included without complications. Care must be used in developing the mesh in $\Delta q^i$, which is strongly dependent on the diagonal elements of the diffusion matrix, e.g.,

$$\Delta q^i \approx (\Delta t g^{ii})^{1/2} \ . \tag{12}$$

Presently, this constrains the dependence of the covariance of each variable to be a (nonlinear) function of that variable, in order to present a straightforward rectangular underlying mesh.

Since integration is inherently a smoothing process [21], fitting data with the short-time probability distribution, effectively using an integral over this epoch, permits the use of coarser meshes than the corresponding stochastic differential equation. For example, the coarser resolution is appropriate, typically required, for numerical solution of the time-dependent path integral. By considering the contributions to the first and second moments conditions on the time, variable meshes can be derived [16]. The time slice essentially is determined by $\Delta t \leq \bar{L}^{-1}$, where $\bar{L}$ is the uniform Lagrangian, respecting ranges giving the most important contributions to the probability distribution $P$. Thus, $\Delta t$ is roughly measured by the diffusion divided by the square of the drift.

Monte Carlo algorithms for path integrals are well known to have extreme difficulty in evolving nonlinear systems with multiple optima [26], but this algorithm does very well on such systems. The PATHINT code was tested against the test problems given in previous one-dimensional systems [16,17]. Two-dimensional runs were tested by using cross products of one-dimensional examples whose analytic solutions are known.

## 5. PATHINT CALCULATION OF DUFFING SYSTEM

While we appreciate that a true calculation of chaos of a stochastic system is quite difficult, here our approach is to simply perform a rigorous numerical calculation of the evolution of the Fokker-Plank equation in its equivalent path-integral representation. Then, we can compare the chaos and non-chaos topologies of the deterministic system, which have been seen above to (barely) be different, to the topologies of the stochastic system. The topology of chaotic systems is an important invariant that often can be used to identify its presence [27].

A mesh of $\Delta t = 0.1$ within ranges of $\{x, y\}$ of $\pm 25$ was reasonable to calculate the evolution of this system for all runs. To be sure of accuracy in the calculations, off-diagonal spreads of meshes were taken as $\pm 5$ increments. This lead to an initial four-dimensional matrix of $159 \times 159 \times 11 \times 11 = 3\,059\,001$ points, which was cut down to a kernel of $2\,954\,961$ points because the off-diagonal points did not cross the boundaries. Reflecting Neumann boundary conditions were imposed by the method of images, consisting of a point image plus a continuous set of images leading to an error function [28], but calculations support the premise that they are unimportant here as the evolving distributions stayed well within the boundaries. Since we have added noise to a set of ordinary differential equations, bringing along additional boundary conditions, we thought it prudent to establish solutions that would be rather independent of any Dirichlet or Neumann boundary conditions.

A Convex 120 supercomputer was used, but there were problems with its C compiler, so gcc version 2.60 was built and used. Runs across several machines, e.g., Suns, Dec workstations, and Crays, checked reproducibility of this compiler on this problem. It required about 12 CPU min to build the kernel and perform the calculation of the next distribution at each $\Delta t$-folding. Data was accumulated after every 5 foldings. The chaos case was run up until $t = 33$, requiring 67.5 CPU hours, and the non-chaos case was run up until $t = 15.5$, requiring 32.0 CPU hours.

The above calculations took over 100 CPU-hours, and this was a self-imposed limit for this first examination of this problem. As demonstrated by the deterministic calculations above, further calculations would have to be performed to enter into the time domain of $t > 50$ after transients have diminished and where chaos truly is strongly exhibited. Furthermore, for $t > 20$, we noticed that there is sufficient diffusion of the system towards the boundaries at $\{x, y\} = \pm 25$, so that these ranges would have to be increased, thereby consuming even more CPU time per $\Delta t$ run. For the purposes of this paper, we just look at epochs where the deterministic calculations show differences between the chaos and non-chaos cases, and compare these to our observations of the stochastic system.

Figs. 3a, 3b and 3c illustrate the stochastic evolution of the phase space for the non-chaos case of $\omega_0 = 1.0$ during a transient period of $t = 12 - 15$; compare to Fig. 2a. Figs. 4a, 4b and 4c illustrate the stochastic evolution of the phase space for the chaos case of $\omega_0 = 0.1$ during a transient period of $t = 12 - 15$; compare to Fig. 2b. The differences in the stochastic cases are practically non-existent, and any differences that are present are likely due to differences in drifts due to changes in the parameters.

## 6. CONCLUSION

We have presented calculations of a two-dimensional time-dependent Duffing system, that has some support for modeling interactions in macroscopic neocortex, that possesses chaotic and non-chaotic regions of activity. We embedded this system in moderate noise and investigated the epochs within the transient periods of these systems to see if there are any apparent differences between the deterministic and stochastic systems.

Future studies taking hundreds of CPU-hours will investigate time epochs beyond the transients to see if moderate noise suppresses chaos in such models of neocortex.

**ACKNOWLEDGEMENT**

RS acknowledges support for this project under NIMH National Research Service Award 1-F32-MH11004-01.

**FIGURE CAPTIONS**

FIG. 1. Phase space plots of $\Phi$ versus $\dot{\Phi}$. (a) The evolution of the phase-space for the non-chaos case of $\omega_0 = 1.0$, after transients have been passed. (b) The evolution of the phase-space for the chaos case of $\omega_0 = 0.1$, after transients have been passed.

FIG. 2. Phase space plots of $\Phi$ versus $\dot{\Phi}$. (a) The evolution of the phase-space for the non-chaos case of $\omega_0 = 1.0$, during the transient period of $t = 12 - 15.5$. (b) The evolution of the phase-space for the chaos case of $\omega_0 = 0.1$, during the transient period of $t = 12 - 15.5$.

FIG. 3. Conditional probability distributions as a function of $\Phi$ versus $\dot{\Phi}$. (a) The stochastic evolution of the phase space for the non-chaos case of $\omega_0 = 1.0$ at $t = 12$. (b) The stochastic evolution of the phase space for the non-chaos case of $\omega_0 = 1.0$ at $t = 13.5$. (c) The stochastic evolution of the phase space for the non-chaos case of $\omega_0 = 1.0$ at $t = 15$.

FIG. 4. Conditional probability distributions as a function of $\Phi$ versus $\dot{\Phi}$. (a) The stochastic evolution of the phase space for the chaos case of $\omega_0 = 0.1$ at $t = 12$. (b) The stochastic evolution of the phase space for the chaos case of $\omega_0 = 0.1$ at $t = 13.5$. (c) The stochastic evolution of the phase space for the chaos case of $\omega_0 = 0.1$ at $t = 15$.

**REFERENCES**

1. P.L. Nunez, *Neocortical Dynamics and Human EEG Rhythms*, Oxford University Press, New York, NY, (1995).

2. A. Fuchs, J.A.S. Kelso, and H. Haken, Phase transitions in the human brain: Spatial mode dynamics, *Int. J. Bifurcation Chaos* **2** (4), 917-939 (1992).

3. V.K. Jirsa, R. Friedrich, H. Haken, and J.A.S. Kelso, A theoretical model of phase transitions in the human brain, *Biol. Cybern.* **71**, 27-35 (1994).

4. L. Ingber and P.L. Nunez, Multiple scales of statistical physics of neocortex: Application to electroencephalography, *Mathl. Comput. Modelling* **13** (7), 83-95 (1990).

5. P.L. Nunez and R. Srinivasan, Implications of recording strategy for estimates of neocortical dynamics with electroencephalography, *Chaos* **3** (2), 257-266 (1993).

6. R. Srinivasan and P.L. Nunez, Neocortical dynamics, EEG standing waves and chaos, in *Nonlinear Dynamical Analysis for the EEG*, (Edited by B.H. Jansen and M. Brandt), pp. 310-355, World Scientific, London, (1993).

7. Y. Ueda, Steady motions exhibited by Duffing's equation: A picture book of regular and chaotic motions, in *New Approaches to Nonlinear Problems in Dynamics*, (Edited by P.J. Holmes), pp. 311-322, SIAM, Philadelphia, (1980).

8. L. Ingber, Statistical mechanics of neocortical interactions. I. Basic formulation, *Physica D* **5**, 83-107 (1982).

9. L. Ingber, Statistical mechanics of neocortical interactions. Dynamics of synaptic modification, *Phys. Rev. A* **28**, 395-416 (1983).

10. L. Ingber, Statistical mechanics of neocortical interactions: A scaling paradigm applied to electroencephalography, *Phys. Rev. A* **44** (6), 4017-4060 (1991).

11. L. Ingber, Statistical mechanics of multiple scales of neocortical interactions, in *Neocortical Dynamics and Human EEG Rhythms*, (Edited by P.L. Nunez), pp. 628-681, Oxford University Press, New York, NY, (1995).

12. L. Ingber, Statistical mechanics of neocortical interactions: Multiple scales of EEG, in *Frontier Science in EEG: Continuous Waveform Analysis (Electroencephal. clin. Neurophysiol. Suppl. 45)*, (Edited by R.M. Dasheiff and D.J. Vincent), pp. 79-112, Elsevier, Amsterdam, (1996).

13.    M. Tabor, *Chaos and Integrability in Nonlinear Dynamics*, Wiley, New York, (1989).

14.    F. Langouche, D. Roekaerts, and E. Tirapegui, *Functional Integration and Semiclassical Expansions*, Reidel, Dordrecht, The Netherlands, (1982).

15.    L. Ingber, Statistical mechanics of nonlinear nonequilibrium financial markets: Applications to optimized trading, *Mathl. Computer Modelling* **23** (7), 101-121 (1996).

16.    M.F. Wehner and W.G. Wolfer, Numerical evaluation of path-integral solutions to Fokker-Planck equations. I., *Phys. Rev. A* **27**, 2663-2670 (1983).

17.    M.F. Wehner and W.G. Wolfer, Numerical evaluation of path-integral solutions to Fokker-Planck equations. II. Restricted stochastic processes, *Phys. Rev. A* **28**, 3003-3011 (1983).

18.    M.F. Wehner and W.G. Wolfer, Numerical evaluation of path integral solutions to Fokker-Planck equations. III. Time and functionally dependent coefficients, *Phys. Rev. A* **35**, 1795-1801 (1987).

19.    L. Ingber, Statistical mechanics of neocortical interactions: Path-integral evolution of short-term memory, *Phys. Rev. E* **49** (5B), 4652-4664 (1994).

20.    L. Ingber and P.L. Nunez, Statistical mechanics of neocortical interactions: High resolution path-integral calculation of short-term memory, *Phys. Rev. E* **51** (5), 5074-5083 (1995).

21.    L. Ingber, Statistical mechanical aids to calculating term structure models, *Phys. Rev. A* **42** (12), 7057-7064 (1990).

22.    L. Ingber, M.F. Wehner, G.M. Jabbour, and T.M. Barnhill, Application of statistical mechanics methodology to term-structure bond-pricing models, *Mathl. Comput. Modelling* **15** (11), 77-98 (1991).

23.    L. Ingber, H. Fujio, and M.F. Wehner, Mathematical comparison of combat computer models to exercise data, *Mathl. Comput. Modelling* **15** (1), 65-90 (1991).

24.    L. Ingber, Statistical mechanics of combat and extensions, in *Toward a Science of Command, Control, and Communications*, (Edited by C. Jones), pp. 117-149, American Institute of Aeronautics and Astronautics, Washington, D.C., (1993).

25.    L. Ingber, Path-integral evolution of multivariate systems with moderate noise, *Phys. Rev. E* **51** (2), 1616-1619 (1995).

26. K. Binder and D. Stauffer, A simple introduction to Monte Carlo simulations and some specialized topics, in *Applications of the Monte Carlo Method in Statistical Physics*, (Edited by K. Binder), pp. 1-36, Springer-Verlag, Berlin, (1985).

27. H.D.I. Abarbanel, R. Brown, J.J. Sidorowich, and L.Sh. Tsimring, The analysis of observed chaotic data in physical systems, *Rev. Mod. Phys.* **65** (4), 1331-1392 (1993).

28. N.S. Goel and N. Richter-Dyn, *Stochastic Models in Biology*, Academic Press, New York, NY, (1974).

**Figure 1a**

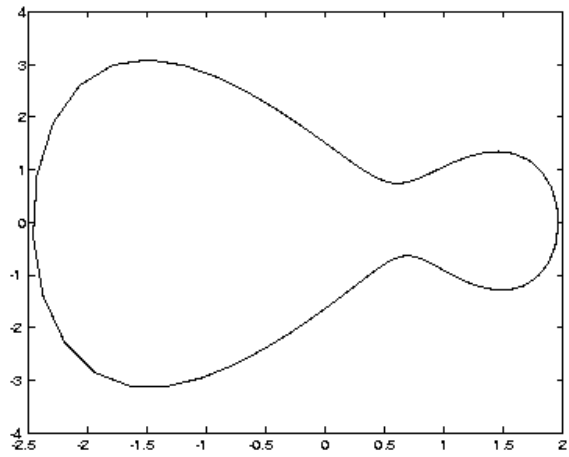

**Figure 1b**

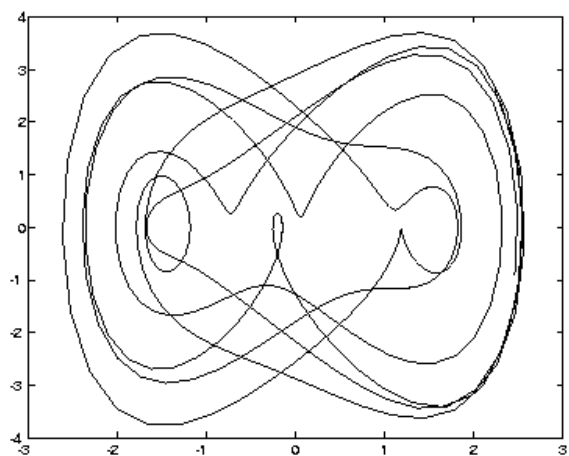

**Figure 2a**

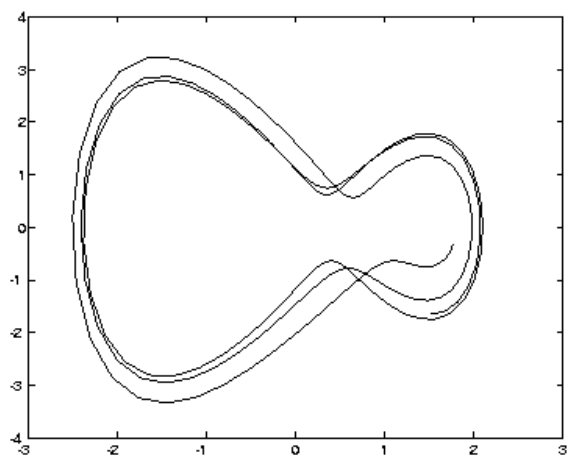

**Figure 2b**

**Figure 2b**

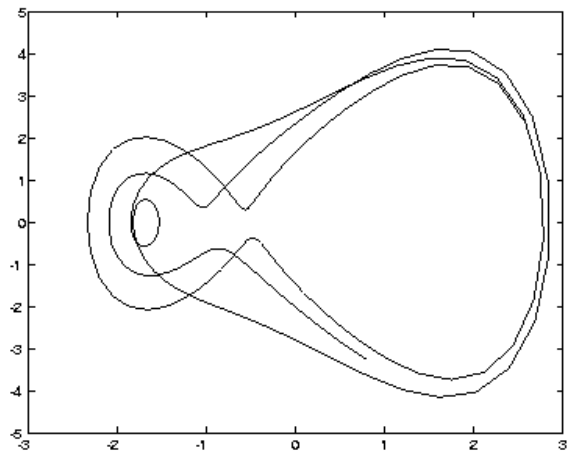

**Figure 3a**

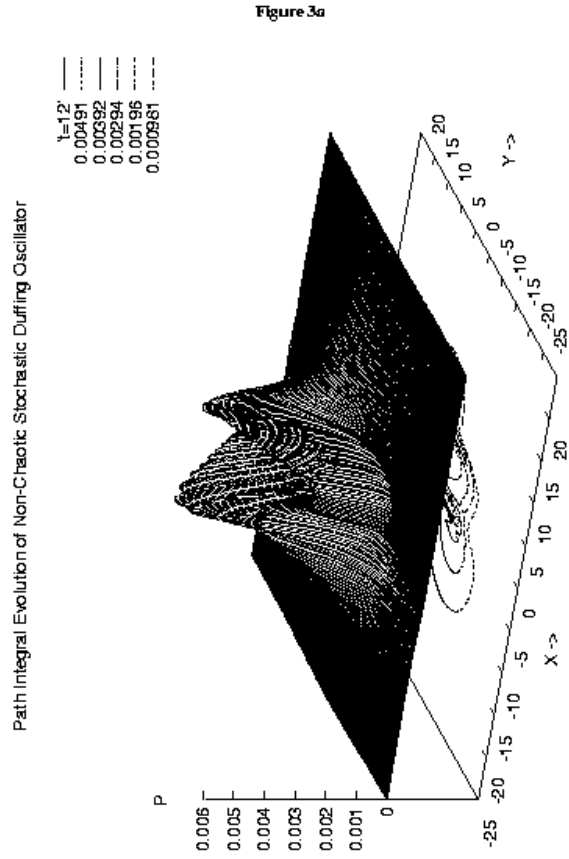

**Figure 3b**

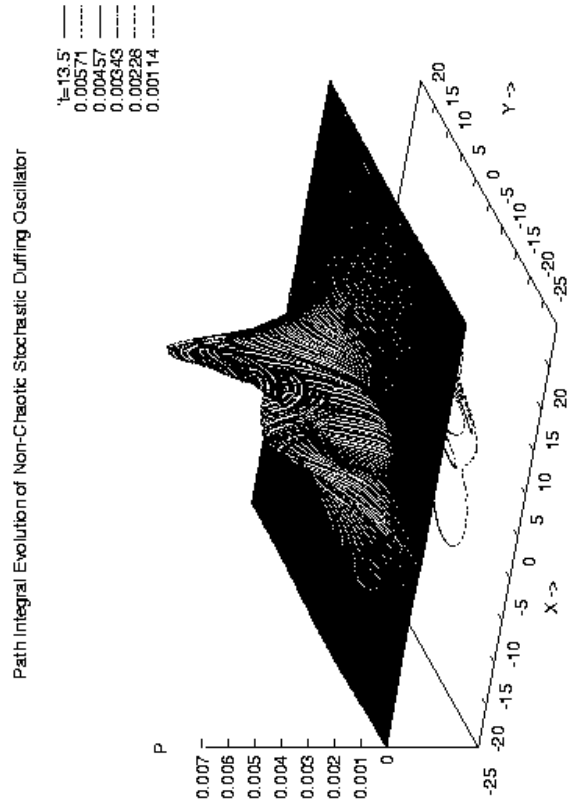

**Figure 3c**

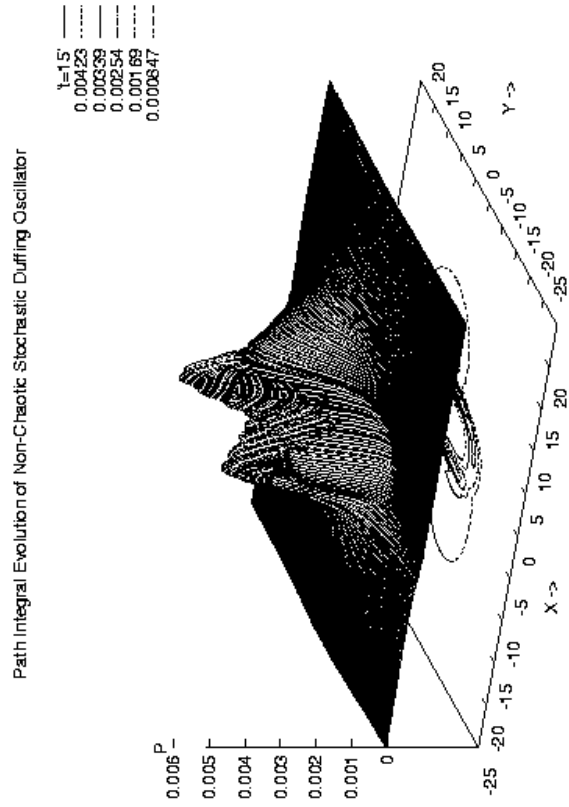

**Figure 4a**

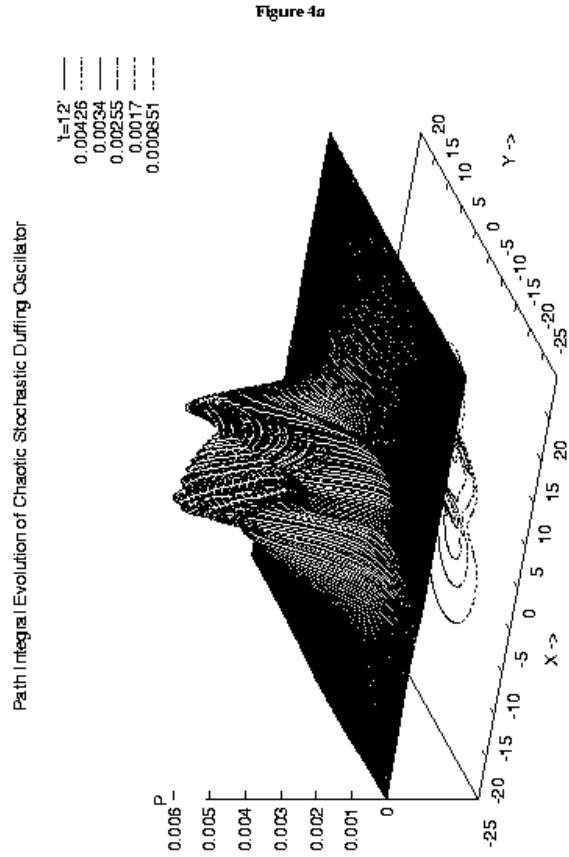

**Figure 4b**

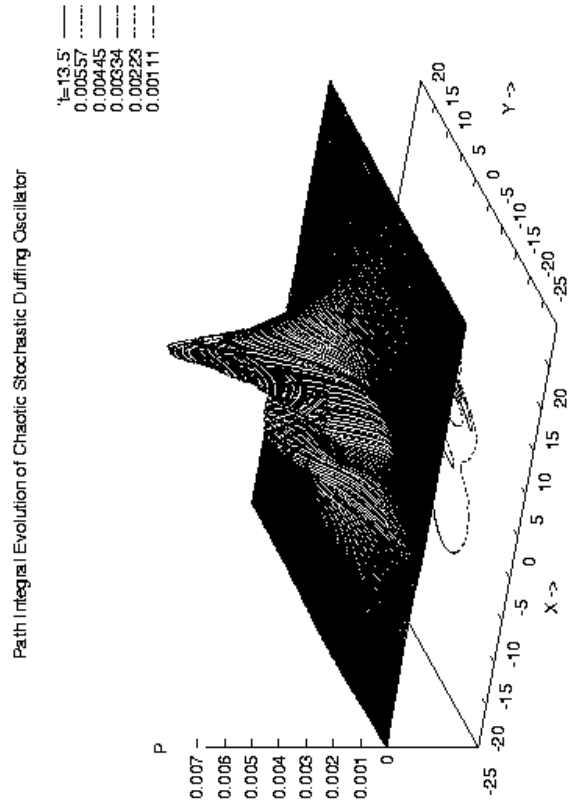

**Figure 4c**

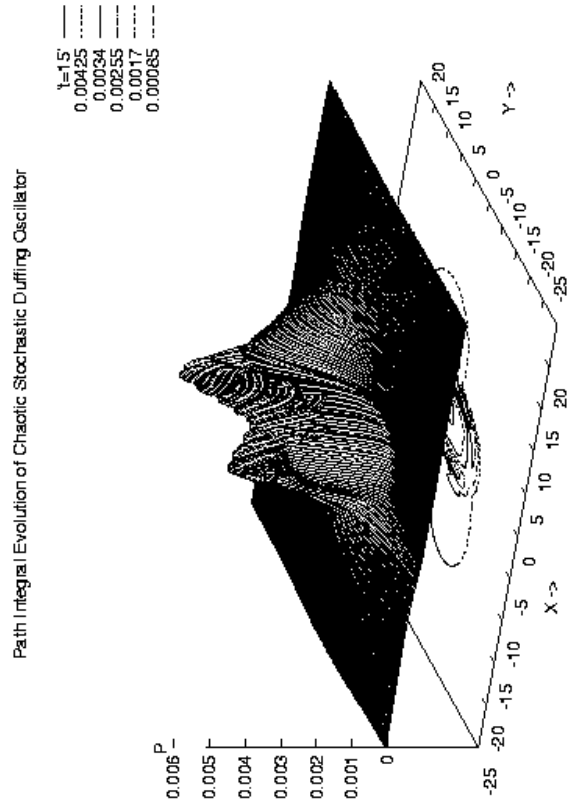